\documentclass[preprint2]{aastex}

\slugcomment{ }

\shorttitle{RFI subtraction}
\shortauthors{Briggs, Bell \& Kesteven}

\begin{document}


\title{Removing radio interference from contaminated astronomical
spectra using an independent reference signal and closure relations}


\author{F. H. Briggs}
\affil{Kapteyn Astronomical Institute, University of Groningen Postbus
800 9700 AV Groningen, The Netherlands; fbriggs@astro.rug.nl}
\author{J. F. Bell \& M. J. Kesteven}
\affil{CSIRO Australia Telescope National Facility, PO Box 76 Epping NSW
1710, Australia; jbell@atnf.csiro.au,~~mkesteve@atnf.csiro.au}




\begin{abstract}
The growing level of radio frequency interference (RFI) is a recognized
problem for research in radio astronomy.  
This paper describes an intuitive but powerful RFI cancellation
technique that is suitable for radio spectroscopy where 
time-averages are recorded. An RFI ``reference
signal,'' is constructed from the cross power spectrum of the signals from the
two polarizations of a reference horn pointed at the source of the RFI
signal.  The RFI signal paths obey simple phase and
amplitude closure relations, which allows computation of the RFI contamination in the
astronomical data and the corrections to be applied to the astronomical spectra.

Since the method is immune to the effects of multipath scattering in both the
astronomy and reference signal channels,
``clean copies'' of the RFI signal are not required.

The method could be generalized (1) to
interferometer arrays, (2) to correct for scattered solar radiation that
causes spectral ``standing waves'' in single-dish spectroscopy, and (3)
to pulsar survey and timing applications where a digital correlator plays
an important role in broadband pulse dedispersion.

Future large radio telescopes, such as the proposed LOFAR and
SKA arrays, will require a high degree of RFI suppression and
could implement the technique proposed here with the benefit
of faster electronics, greater digital precision and higher
data rates.
\end{abstract}




\section{Introduction}

The growing level of radio frequency interference (RFI) is a recognized
problem for research in radio astronomy.  Fortunately, the technological
advances that are giving rise to the increasing background of radiation --
through increased telecommunications and wide-spread
use of high speed electronics --
are also providing some of the tools necessary for separating astronomical
signals from undesirable RFI contamination.  New radio telescopes will
necessarily have RFI suppression, excision and cancellation algorithms
intrinsic to their designs.  No one technical solution will make radio
observations immune to interference; successful mitigation is most likely to
be a hierarchical or progressive approach throughout the telescope,
combining new instrumentation and algorithms for signal conditioning and
processing \cite{eb00,ell99}.

Techniques from the communication industry that are finding application in
radio astronomy experiments include (1) adaptive beam forming with array
telescopes that steer nulls of the instrument reception pattern in the
directions of sources of RFI \cite{hgjs98,eh00,lvb00,sh00,ksb+00}, (2)
parametric signal modeling techniques, where the RFI signal is received and
decoded to obtain a high signal-to-noise ratio (SNR) reference signal for
subtraction from the astronomical data \cite{ebb00,lv99,lv99a}, and (3)
adaptive filtering using a reference horn to obtain a high SNR copy of the
RFI for real-time cancellation from the signal path ahead of the standard
radio astronomical backend processors \cite{bb98}.

\begin{figure}[htbp]
\epsscale{.75}
\plotone{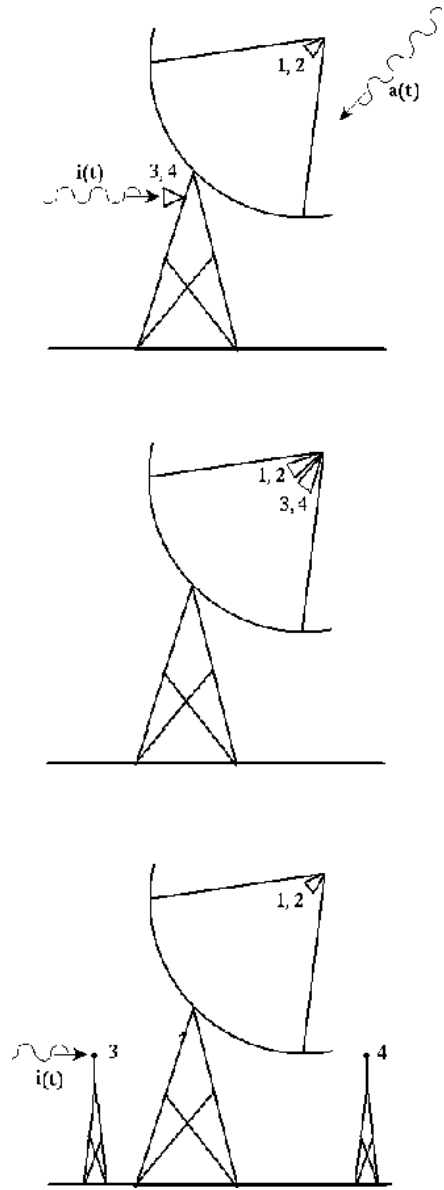}
\caption{Three configurations applicable to the analysis
in this paper. {\it Top} The four basebands are recorded from
the 2 polarizations of the Parkes telescope feed and 2
polarizations of a reference horn directed at the RFI source.
{\it Center} The four basebands correspond to 2 polarizations
from each of two Parkes feeds. {\it Bottom} A proposed system
for optimal application of the RFI subtraction technique
described here.
}
\label{telescopes.fig}
\end{figure}

This paper describes an intuitive but powerful RFI cancellation
technique that is suitable for radio spectroscopy where 
time-averages are recorded. 
The method requires computation of cross power spectra between
the RFI contaminated astronomical signals and high signal-to-noise
ratio RFI ``reference signals'' obtained from a receiving system that senses
the RFI but not the astronomical signal. The correction term that
removes the unwanted RFI is computed from closure relations obeyed
by the RFI signal.  The test applications reported here
derived the  reference signal either from a separate horn antenna
aimed at the RFI source or from a second feed horn at the focus of
the Parkes telescope, as illustrated in Fig~\ref{telescopes.fig}.
For these experiments, we recorded digitally sampled baseband signals
from two polarizations for both the reference and astronomy feeds, and
then we performed the cross correlations in software off-line.
However, the method could use correlation
spectrometers of the sort already in use at radio observatories. 
With minor design enhancements, future generation correlators
could incorporate this technique with the 
additional benefit of the faster electronics, greater digital precision,
and higher data rates that technological advance promises.
 
There are a number of advantages to performing the RFI in a
``post-correlation'' stage.  Foremost is that the RFI subtraction
remains an option in the data reduction path, rather than
a commitment made online and permanently.
Furthermore, the  correlation method
is effectively a coherent subtraction, since the correlation
functions retain the information describing relative phase between
the RFI entering in the astronomy data stream and the RFI entering
the reference antenna. As we show in this paper, this means that the RFI  
noise power is largely subtracted, leaving only the usual components
of system noise.

This paper provides a description of the post-correlation RFI cancellation
technique and illustrates its success with data from the Parkes telescope.
A mathematical overview shows (1) why unknown multipaths do not cause the
algorithm to break down, (2) how to simply construct a suitable RFI
reference spectrum, and (3) how to build an inverse filter to obtain
immunity to low signal levels at frequencies that suffer destructive
interference by multipathing in the reference horn signal path.

\section{Mathematical description of the method}

In our mathematical model, 
we make the assumption that the RFI source emits a single signal
$i(t)$.  (At the RFI source, the signal from a single power amplifier 
feeds an antenna of
unknown, but irrelevant, polarization.)  The RFI that appears differently in
the recorded data channels at Parkes has experienced scatterings with
different path lengths and amplitudes, so that the received signals are
linear sums of time delayed versions of the original  broadcast $i(t)$.

In fact, the model is applicable to multiple interferers within
the spectrometer band, provided they do not overlap in frequency.
 
Consider for the moment a single interferer, which propagates through
the four signal paths $s_1$, $s_2$, $s_3$, and $s_4$ that will be 
processed:  there are
two astronomical channels, $s_1(t)$ and $s_2(t)$, which convey the voltages
$a_A(t)$ and $a_B(t)$ from the celestial sources for the two independent
polarizations from the Parkes Telescope receiver along with contamination
from the RFI signal $i(t)$.  Radiation from astronomical sources may
be partially polarized, causing $a_A(t)$ and $a_B(t)$ to be correlated to
some degree. 
The two reference channels carry $s_3(t)$ and $s_4(t)$,
containing the representations of $i(t)$ but negligible signal from the
celestial sources.  For example, the measured signal in channel 1
comprises channel noise $n_1(t)$, 
plus the convolutions of the impulse responses for
the multiple scattering paths for the interference signal and the true
astronomy signal: $s_1(t)= H_A(t)\bigstar a_A(t) + H_1(t)\bigstar i(t)+n_1(t)$. Here
$H_A(t)$ and $H_1(t)$ are impulse responses for the astronomy signal path
and interference respectively, and $\bigstar$ is the convolution operator.

For many purposes, an intuitive picture of the multipathing results
from considering the scattering sites to be achromatic mirror-like
scatterers each with relative effective areas $G_{A,j}$ and $G_{1,k}$, and
attaching the path delay to each separate version of the astronomy and RFI
signals, $s_1(t)=\sum_{j} G_{A,j} a_A(t-\tau_{1,j}) + \sum_{k} G_{1,k}
i(t-\tau_{1,k}) + n_1(t)$. The time delays $\tau$ are determined by the
different path lengths $L_{1,j}$ to give $\tau_{1,j}=L_{1,j}/c$, where $c$
is the speed of light.

All the signal paths are vulnerable to corruption by
stochastic noise.
The noise terms  $n_1(t)$, $n_2(t)$, $n_3(t)$, and $n_4(t)$
should ideally be uncorrelated among the
different data paths.  Unfortunately, in real astronomical systems, there
is likely to be low level coupling between the two orthogonal polarizations of
a feed horn or common stray radiation pickup from spillover that will
make a weakly correlated noise floor in some of the cross power spectra.
This will form a systematic limitation to the accuracy of the subtraction.

In this experiment, the goal is to explore the usefulness of
cross-correlation spectra to correct for the effects of RFI in
time-averaged spectra.  
These  spectra are products of scaled sums of the
of the Fourier transforms of the astronomy signals, the RFI and noise.
In the tests with real data in section~\ref{subtraction_section}, we
compute estimates of the complex spectra
\begin{eqnarray}
S_1(f) & = & g_A A_A + g_1 I + N_1 \nonumber \\
S_2(f) & = & g_B A_B + g_2 I + N_2 \nonumber \\
S_3(f) & = &   g_3 I + N_3 \nonumber \\
S_4(f) & = &   g_4 I + N_4 
\label{voltage_spectra.eqn}  
\end{eqnarray}
from Fourier transforms of finite length
time series of discrete samples of the four signals. The transforms 
contain  contributions from the celestial sources $A_A(f)$ and
$A_B(f)$, the RFI $I(f)$, and the noise in each channel $N_i(f)$,
modulated by the associated complex voltage gains, which are
the Fourier transforms of the impulse responses $H(t)$. The gains for
each channel separate into dependencies on (1) the path delay $L/c$,
which appears in a frequency dependent phase term,
according to the Shift Theorem of Fourier transforms, and (2) a
possible additional frequency dependence $g(f)$ of each delay path:
\begin{equation}
g_1(f) = \sum  g_{1,k}(f)e^{i2\pi fL_{1,k}/c}
\end{equation}
These complex gain and delay factors are sufficiently
general to include complicated scatterers and propagation through
dispersive and lossy media.

The real power spectra for the four data channels have the following
form, once terms that average toward zero are omitted and the
complex gains are assumed to be constant over the time span for which 
the spectra are computed:
\begin{eqnarray}
P_1(f) & = & <S_1S_1^*> \nonumber \\
& = & |g_A|^2 <|A_A|^2> \nonumber \\
& & ~ ~ + |g_1|^2 <|I|^2> + <|N_1|^2> \nonumber \\
P_2(f) & = & <S_2S_2^*> \nonumber \\
& = & |g_B|^2 <|A_B|^2> \nonumber \\
& & ~ ~ + |g_2|^2 <|I|^2> + <|N_2|^2> \nonumber \\
P_3(f) & = & <S_3S_3^*>  \nonumber \\
& = &  |g_3|^2 <|I|^2> + <|N_3|^2>\nonumber \\
P_4(f) & = & <S_4S_4^*>  \nonumber \\
& = &  |g_4|^2 <|I|^2> + <|N_4|^2>
\label{acspectra.eqn}
\end{eqnarray} 
We use the superscript symbol $^*$ to represent complex conjugation and
the $<...>$ notation to signify averages over an integration
time $t_{int}$; in the 
tests we describe in section~\ref{subtraction_section},
we find the method is effective for $t_{int}$ as long as $\sim$1~sec.
We adopt a normalization where the power levels 
$<|A_A|^2> \approx  <|I|^2>\approx<|N_1|^2>$ so that,
for example, in data channel 1
the signal to noise ratio $SNR_1\sim |g_A|^2$, and the
interference to noise ratio $INR_1\sim |g_1|^2$.

The goal of the RFI cancellation will be to form estimates of
the $|g_1|^2<|I|^2>$ and $|g_2|^2<|I|^2>$ terms and then subtract
them from $P_1(f)$ and $P_2(f)$, while leaving the astronomical signal
(and noise) behind.

 This discussion has
assumed that the complex gain terms are constant over the integration time
$t_{int}$, allowing us to separate the interference from the gain 
in expressions such as:
\begin{equation}
<g_1Ig_1^*I^*> = |g_1|^2<|I|^2>
\label{time_dep.eqn}
\end{equation}
In anticipation of the discussion of section~\ref{subtraction_section}, we
note that this assumption will fail for extended integration times
since the scattering paths that lead the RFI signal to the telescope
feed will change,
resulting in loss of precision in the cancellation scheme  and 
and leading to substantial residuals in the corrected spectra.

The complex cross power spectra for all combinations of the four
data channels have the following form when  the leading
 contributions are retained:
\begin{eqnarray}
C_{12}(f) & = & <S_1S_2^*> \nonumber \\ 
& = & g_Ag_B^*  <A_AA_B^*> \nonumber \\
& & ~ ~ ~+ g_1g_2^* <|I|^2> \nonumber \\
& & ~ ~ ~+ g_1<IN_2^*> + g_2^*<N_1I^*> \nonumber \\
& & ~ ~ ~+ <N_1N_2^*>\nonumber \\
C_{ij}(f) & = &  <S_iS_j^*> \nonumber \\ 
& = & ~ ~ ~ g_ig_j^* <|I|^2> \nonumber \\
& & ~ ~ ~+ g_i<IN_j^*> + g_j^*<N_iI^*> \nonumber \\
& & ~ ~ ~+ <N_iN_j^*>\nonumber \\
&   & ~ ~ ~ ~   \rm{~ for~ } i\ne j, ~ ~ j>2 
\label{ccspectra.eqn}
\end{eqnarray}

In order to cancel the dominant RFI terms in the power spectra
Eqns~\ref{acspectra.eqn}, we need to compute quantities of the
form $|g_1|^2 <|I|^2>$, which can be subtracted from the
measured $<|S_1|^2>$. One possibility would be  a combination of
auto and cross power spectra of the form
\begin{eqnarray}
|g_1|^2 <|I|^2> & = & \frac{g_1g_3^*g_1^*g_3}{g_3^*g_3}<|I|^2> 
    \nonumber \\ 
& \approx & \frac{|C_{13}|^2}{<|S_3|^2>}  \\ 
  & \approx & \frac{|C_{13}|^2}{|g_1|^2 <|I|^2> + <|N_3|^2>}  \nonumber
\label{bad1.eqn}
\end{eqnarray}
However, the problem encountered with this approach is that $<|S_3|^2>$
is biased by the term $<|N_3|^2>$, which averages over time to
the total power spectrum of the noise in data channel 3. This
might be compensated by calibrating the noise spectrum, in order to 
improve the estimate of $|g_3|^2<|I|^2>=<|S_3|^2>-<|N_3|^2>$ 
to use in the denominator of Eqn~\ref{bad1.eqn}.

An alternate combination that avoids this bias
forms the estimate of $|g_1|^2 <|I|^2>$
from three cross power spectra, in which the
noise terms have the form $<N_iN_j^*>$ and  $<IN_j^*>$ which 
average toward  zero as the integration time increases: 
\begin{eqnarray}
|g_1|^2 <|I|^2> & = & \frac{g_1g_3^*g_1^*g_4}{g_3^*g_4}<|I|^2> 
    \nonumber \\ & \approx & \frac{C_{13}C_{14}^*}{C_{34}^*} \label{corr1.eqn}\\
|g_2|^2 <|I|^2> & = & \frac{g_2g_3^*g_2^*g_4}{g_3^*g_4}<|I|^2> 
   \nonumber \\ & \approx & \frac{C_{23}C_{24}^*}{C_{34}^*} \label{corr2.eqn} \\
g_1g_2^* <|I|^2> & = & \frac{g_1g_4^*g_2^*g_3}{g_3g_4^*}<|I|^2> 
    \nonumber \\&  \approx  &\frac{C_{14}C_{23}^*}{C_{34}} \label{corr3.eqn}  
\end{eqnarray}
The expressions involving the measured cross power spectra are
``approximate,'' since the
cross power spectra result from finite integrations, and the noise
terms will  limit the precision of the cancellation. 

The occurrence of the $C_{34}$ term in the denominator for
Eqns~\ref{corr1.eqn}-\ref{corr3.eqn} indicates there will be a problem in
implementing a correction scheme in frequency ranges where $C_{34}$ 
becomes small or zero. In many
situations when the signal to noise ratio for the spectra in the numerator
is very high, the cross power spectra in the numerator will also be small or
zero whenever $C_{34}$ is small, so that divergence will be canceled. A simple
means  to avoid division by small numbers in
the presence of noise in this kind of situation is to create a filter of the
following form:
\begin{eqnarray}
|g_1|^2 <|I|^2> & \approx &  
        \frac{C_{13}C_{14}^*C_{34}}{\psi(f)+C_{34}C_{34}^*} \label{wein1.eqn}\\
|g_2|^2 <|I|^2>& \approx &     \frac{C_{23}C_{24}^*C_{34}}{\psi(f)+C_{34}C_{34}^*}\label{wein2.eqn}\\
g_1g_2^* <|I|^2> & \approx & 
        \frac{C_{14}C_{23}^*C_{34}^*}{\psi(f)+C_{34}C_{34}^*}
 \label{wein3.eqn}
\end{eqnarray}
where $\psi(f)$ is the square of the power spectrum of the noise 
present in $C_{34}(f)$.  
Whenever $\psi(f)$ becomes small compared to $C_{34}(f)$, the expressions
Eqns~\ref{wein1.eqn}-\ref{wein3.eqn} revert to 
Eqns~\ref{corr1.eqn}-\ref{corr3.eqn}. 
When the noise exceeds the signal power in $C_{34}$, the computed
correction tends to zero. 
In practice, during the  
test described here, a constant  $\psi_0$ 
was used in place of $\psi(f)$.
 Alternatively, division by zero can
be avoided by testing the amplitude of $C_{34}(f)$ 
for significance above a noise threshold  and setting the correction
to zero when the significance criterion is not met.

These corrections for the the auto correlation spectra
(Eqns~\ref{corr1.eqn} and \ref{corr2.eqn}) are expected
to be real valued. Therefore a logical test for the accuracy of the
correction is that phases computed for the frequencies of strong RFI signal
should be close to zero. In fact, this is a statement of phase closure. 
Note that the denominators of 
Eqns~\ref{wein1.eqn}-\ref{wein3.eqn} are purely real, and the
numerators, such as $C_{13}C_{14}^*C_{34}$, form logical triangles
for computing closure phases.

An amplitude closure relation 
\begin{equation}
\frac{C_{13}C_{24}}{C_{23}C_{14}}=\frac{g_1g_3^*g_2g_4^*}{g_2g_3^*g_1g_4^*}=1
\label{amp_closure.eqn}
\end{equation}
can also be constructed and tested. It too will suffer from divergence of
the quotient in frequency ranges where the cross power spectra
are noisy and $C_{23}$ and $C_{14}$ 
have small amplitude. Here we avoid including $C_{12}$, which typically
has  significant
cross correlated power in addition to the RFI signal, due either to
polarized celestial flux or cross talk between the channels. 

\onecolumn
\section{Noise and the Accuracy of the Cancellation}
\label{accuracy_section}

In this section we assess the importance of the Interference-to-Noise
ratio.  First we expand the autocorrelation spectra, keeping all
cross terms, including those that average toward zero.  Then
$P_1(f)$ becomes
\begin{eqnarray}
P_1(f) & = &  |g_A|^2 <|A_A|^2> + ~ |g_1|^2 <|I|^2>
+~2{\rm Re}\left[g_1g_A^*<IA^*>+~g_1<IN_1^*>\right] \nonumber \\
 & & ~ ~ ~
+~2{\rm Re}\left[ g_A<AN_1^*>\right]~
 + <|N_1|^2> 
\label{acspect_exp.eqn}
\end{eqnarray}
The complex correction spectra were described by 
equations~\ref{corr1.eqn}, \ref{corr2.eqn} and \ref{corr3.eqn}.  Including the
cross terms and  noting that when 
\begin{equation}
g_3 \rm {~ and~} g_4 \gg g_1 \rm {~ and~} g_2  \label{cond_1.eqn}
\end{equation}
and the interference power to noise power ratios
\begin{eqnarray}
INR_3 & = & \frac{|g_3|^2<|I|^2>}{<|N_3|^2>} ~~\approx ~ |g_3|^2 ~~ \gg ~ 1 \nonumber \\
INR_4 & = & \frac{|g_4|^2<|I|^2>}{<|N_4|^2>} ~~\approx ~ |g_4|^2 ~~ \gg ~ 1  \label{cond_2.eqn}
\end{eqnarray}
then the correction $CX_1$ for $P_1(f)$  becomes:
\begin{eqnarray}
CX_1 & \approx & \frac{C_{13}C_{14}^*}{C_{34}^*} \nonumber \\
 &\approx & |g_1|^2 <|I|^2> +
 ~ 2{\rm Re}\left[g_1g_A^*<IA^*>+~g_1<IN_1^*>\right]  \nonumber \\
 & & ~ + ~ \frac{g_Ag_1^*}{g_3^*}<AN_3^*> + ~ \frac{g_A^*g_1}{g_4^*}<A^*N_4>
  \nonumber \\
 & & ~ + ~ \frac{g_1^*}{g_3^*}<N_1N_3^*> + ~ \frac{g_1}{g_4}<N_1^*N_4> 
 - ~ \frac{|g_1|^2}{g_3^*g_4}<N_3^*N_4>  
\label{corr1x.eqn}
\end{eqnarray}
The terms in Eqn~\ref{corr1x.eqn} involving $I$ all appear in the
power spectrum of Eqn~\ref{acspect_exp.eqn}, so that
application of this correction $CX_1$ to $P_1(f)$ leads to
a result with no residual contamination by the RFI: 
\begin{eqnarray}
P_1(f)~- ~ CX_1 & = &  |g_A|^2 <|A_A|^2> + ~  <|N_1|^2> \nonumber \\
 & & ~ ~ ~ + ~2{\rm Re}\left[ g_A<AN_1^*>\right]~ 
 - ~ \frac{g_Ag_1^*}{g_3^*}<AN_3^*> - ~ \frac{g_A^*g_1}{g_4^*}<A^*N_4>  \nonumber \\
 & & ~ ~ ~ - ~ \frac{g_1^*}{g_3^*}<N_1N_3^*> - ~ \frac{g_1}{g_4}<N_1^*N_4> 
 + ~ \frac{|g_1|^2}{g_3^*g_4}<N_3^*N_4>  
\label{acspect_corr.eqn}
\end{eqnarray}
The complete cancellation of the RFI terms 
is consistent with the concept that post
correlation subtraction is equivalent to coherent subtraction of
the RFI electric field $i(t)$ in the time domain, 
which should leave no trace of the RFI signal nor an increase in noise power.
Provided the $INR$ for the reference horn channels is substantially
greater than the $INR$ for the Parkes feed channels, the noise terms
due to $N_3$ and $N_4$ will be smaller by a factor of order 
$\beta\sim\sqrt{INR_3/INR_1}\approx |g_3|/|g_1|$ than the normal noise contributions
arising from $N_1$ plus the astronomical signal power. The terms 
in Eqn~\ref{acspect_corr.eqn}
other than $|g_A|^2 <|A_A|^2>$ and  $<|N_1|^2>$, such as 
$(g_A^*g_1/g_4^*)<A^*N_4>\propto |g_A|\beta^{-1} t^{-1/2}$ 
and
$(g_1^*/g_3^*)<N_1N_3^*>\propto \beta^{-1} t^{-1/2}$,
 average toward zero
(in inverse proportion to the square root of the integration time),
provided the noise in the signal channels is independent.
The next higher order terms, which are not included in 
 Eqn~\ref{acspect_corr.eqn},
have dependencies such as $<IN_1^*><N_1^*N_4>/g_4<|I|^2>\propto t^{-1}$
and $g_1^*<IN_3^*><I*N_1>/g_3^*<|I|^2>\propto \beta^{-1}t^{-1}$,
which converge toward zero faster than the dominant noise terms.
\twocolumn

\newpage
\begin{figure}[htbp]
\epsscale{1.0}
\plotone{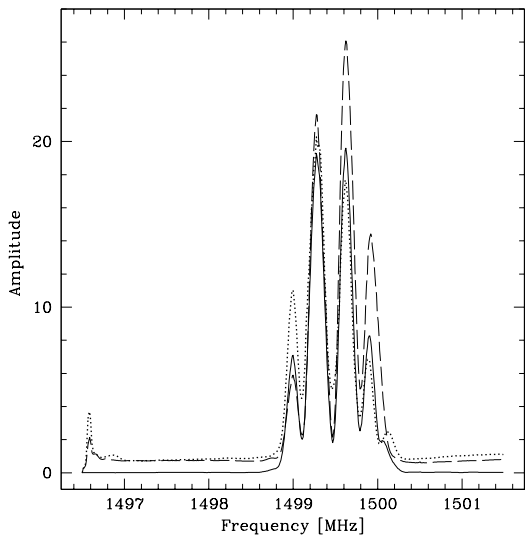}
\caption{Scan averaged RFI spectra measured with the reference horn for scan
SRT00502. Solid line is the cross power spectrum $C_{34}(f)$ defined in
Eqn~\ref{ccspectra.eqn}.  Dashed and dotted lines indicate the
autocorrelation spectra $P_3(f)$ and $P_4(f)$ in Eqn~\ref{acspectra.eqn}.
The spectra have been passband calibrated using approximate gain curves for
the 5 MHz filters. There are 512 frequency channels covering a 5 MHz band.}
\label{rfi_spectra.fig}
\end{figure}

\section{The test data}
\label{subtraction_section}

 The astronomical dataset used in testing these algorithms is a dual linear
polarization data stream from the CSIRO ATNF 64m telescope at Parkes in
Australia. One configuration has two polarizations from the central beam of
the the Parkes multibeam receiver \cite{swb+96} and two polarizations from a
reference horn aimed at an interfering source \cite{bhw+00}.  A second
configuration uses both polarizations from two beams of the Parkes Multibeam
system, which are directed at slightly different areas of the sky.  The
center frequency of the datasets was 1499~MHz in each case.

The datasets we used are labeled
SRT00501, SRT00502 and SRT00601 \cite{bhw+00}. The main interfering
source is a NSW Government digital point-to-point microwave link.
Examples of the time-averaged spectrum for the RFI signal are
shown in Fig~\ref{rfi_spectra.fig}.  Further
details are available in the ACA databases \cite{aca98,sar99}. The 4 signals
were downconverted to base band and passed through 5~MHz low pass
filters. Each signal was then digitized with 2-bit precision at a
20MHz sampling rate to achieve a factor 2 oversampling and recorded using the
CPSR recorder \cite{vbb+99}. 
 
The data processing for these tests
simulates a radio astronomy backend by computing power spectra
and cross \-power spectra in software. 
The sampled voltages are treated in 8192 sample blocks (410$\mu$sec
durations).  Fourier transformation of each block yields a spectrum
of 4096 independent complex coefficients.

\begin{figure}[htbp]
\epsscale{1.0}
\plotone{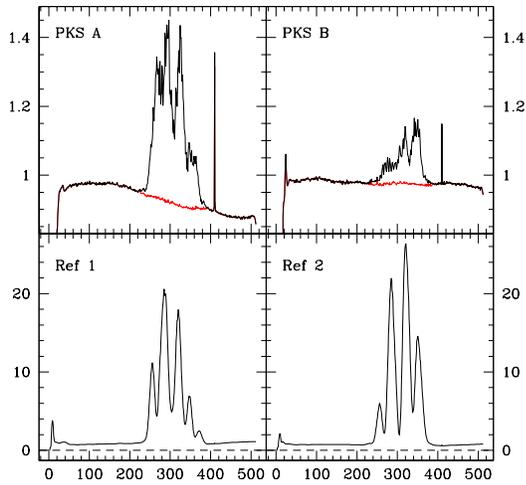}
\caption{The Power Spectra PKS~A$=P_{1}(f)$, PKS~B$=P_{2}(f)$, 
Ref~1$=P_{3}(f)$, and Ref~2$=P_{4}(f)$
for scan SRT00502.  These spectra are the averages of $\sim$25 seconds of
data. A passband calibration has been applied to compensate for the gain
dependence of the 5 MHz band limiting filters. The upper panels
show the spectra both before and after cancellation.}
\label{raw_auto_power.fig}
\end{figure}

Examples of 25 second averages of the power spectra are shown in 
Fig~\ref{raw_auto_power.fig}.
The RFI spectrum is significantly different in the spectra for
all four data channels.
Since the data are 2$\times$ oversampled, we kept the total power
and cross power spectra of the lower 2048 complex Fourier coefficients.
The RFI subtraction steps were performed on these spectra, as
illustrated below, and subsequently the
spectra were block averaged by 4 to keep power spectra of length 512 
spectral channels for further calibration and display at 9.8~kHz resolution.

\begin{figure}[htbp]
\epsscale{1.0}
\plotone{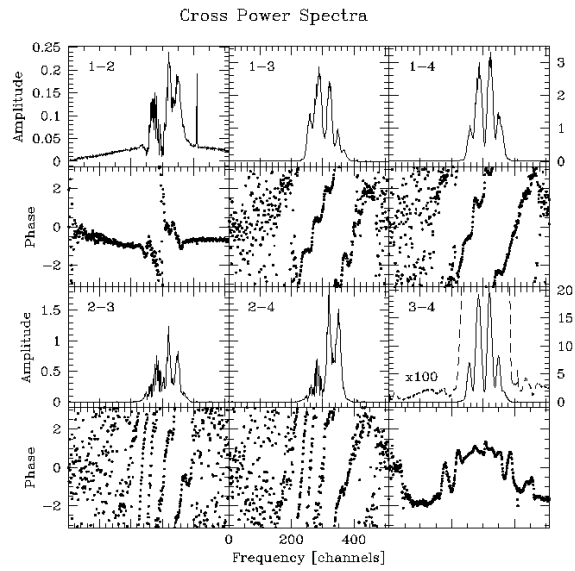}
\caption{The Cross Power Spectra $C_{12}(f), C_{13}(f), C_{14}(f),
C_{23}(f), C_{24}(f),$ and $C_{34}(f)$ for scan SRT00502.  These spectra are
the averages of $\sim$25 seconds of data. The spectrum $C_{34}(f)$ at lower
right is also plotted with a rescaling of a factor of 100 in order to
display the noise level away from the frequencies containing strong RFI.}
\label{raw_cross_power.fig}
\end{figure}

Examples of the cross power spectra are shown in 
Fig~\ref{raw_cross_power.fig}. Both $C_{12}$ and $C_{34}$ have broadband
correlated power, which is clear in the integrated
spectra both as a significant non-zero amplitude and
as a well defined trend  in  phase across the band. The
phase gradients that are clearly visible in the $C_{13}(f), C_{14}(f),
C_{23}(f)$, and $C_{24}(f)$ spectra indicate differential path delays.
The delay for $C_{13}(f), C_{14}(f)$ and for the channel numbers above
300 in $C_{23}(f)$, and $C_{24}(f)$ amounts to 15 time steps (0.75$\mu$sec).
The steeper phase gradient for channels below 300 in 
$C_{23}(f)$, and $C_{24}(f)$ imply path delays of 60 time steps $=3\mu$sec
or a path length of $\sim$900m.  Apparently, the frequency dependent
reception pattern of the
far sidelobes of the Parkes dish, coupled with complicated scatterers
in the field, can lead to quite complex spectral dependence for the RFI.
A strength of our method is the simplicity with which it handles this
complex multipathing. 

\begin{figure}[htbp]
\epsscale{1.0}
\plotone{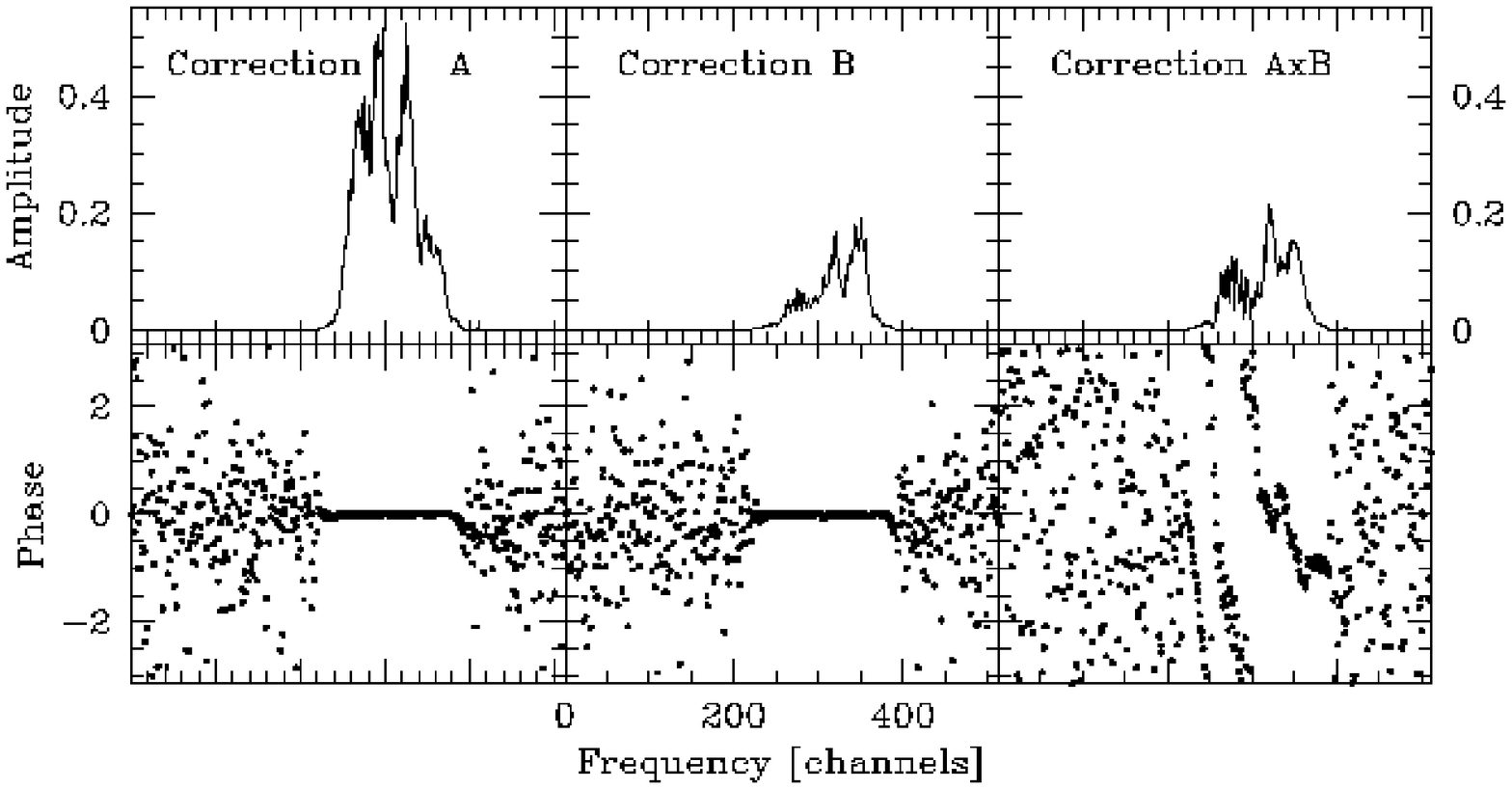}
\caption{The Complex Correction Spectra derived from
Eqns~\ref{wein1.eqn}-\ref{wein3.eqn} for scan SRT00502 for application
to the Parkes spectra A, B, and A$\times$B.  These plots show
the averages of $\sim$25 seconds of data.  The RFI subtraction was actually
performed on 82 msec averages. A \& B indicate the 2 polarizations.}
\label{error_spectra.fig}
\end{figure}

Examples of the complex correction spectra defined in
Eqns~\ref{corr1.eqn}-\ref{corr3.eqn} are shown in 
Fig~\ref{error_spectra.fig}. The spectra for Correction A and
Correction B have zero phase over the spectral range where the signal
is significantly non-zero, confirming that phase closure applies.  

\begin{figure}[htbp]
\epsscale{1.0}
\plotone{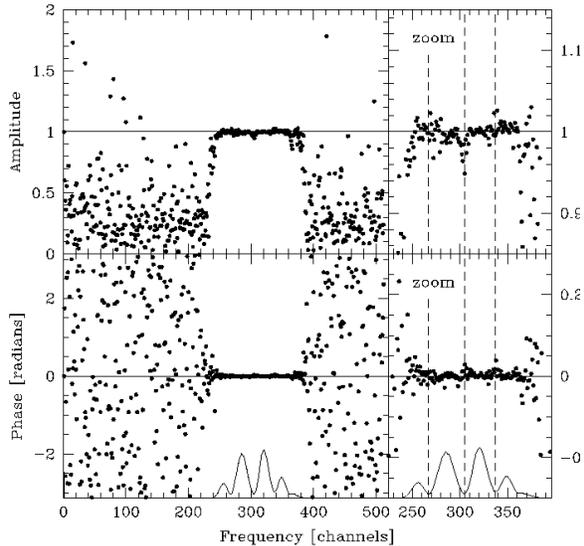}
\caption{The complex closure quantity ${C_{13}C_{24}}/{C_{23}C_{14}}$
averaged for scan SRT00502.
{\it Left}: full spectrum. {\it Right}: expanded scales. Reduced
scale plots of the average RFI cross power spectrum from
Fig~\ref{rfi_spectra.fig} are drawn in the bottom panels for comparison.
}
\label{closure.fig}
\end{figure}

Fig~\ref{closure.fig}
shows averages of the amplitude closure quantity from 
Eqn~\ref{amp_closure.eqn}, plotted as
amplitude and phase across the spectrum. There is
deterioration in the closure relation when the RFI signal is low,
as expected.

Fig~\ref{rfi_spectra.fig} compares three time-averaged power spectra
representing the RFI signal sensed by the reference horn: power spectra for
each of the two polarizations and one cross power spectrum.  For use in
correction schemes, the cross power spectrum $C_{34}(f)$ 
has the advantage that
it is not contaminated by the positive bias of the receiver noise total
power that is seen in the autocorrelation power spectra. In practice, the
low level correlated signal in $C_{34}(f)$, 
due to actual cross correlated broadband power
or due to correlated
quantization noise, may limit the accuracy of the cross power
spectrum as an estimate of the RFI.
\begin{figure}[htbp]
\epsscale{1.0}
\plotone{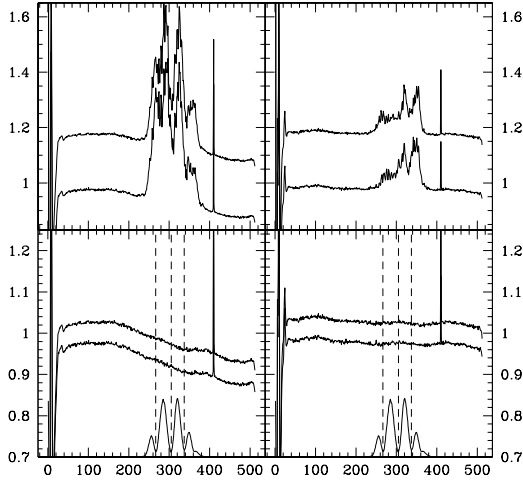}
\caption{Comparison of two scans: SRT00501 and SRT00502.
{\it Upper Panels:} Uncorrected power spectra for the two Parkes
polarizations. Spectra for SRT00501 are displaced vertically by
0.2 in amplitude.
{\it Lower Panels:} Corrected power spectra for the two Parkes
polarizations. Spectra for SRT00501 are displaced vertically by
0.05 in amplitude. Reduced scale copies of the RFI cross power spectrum are 
included for reference with vertical dashed lines to indicate the minima
in the RFI spectrum.
}
\label{result.fig}
\end{figure}

\begin{figure}[htbp]
\epsscale{1.0}
\caption{Dynamic Power Spectra over 564 time steps of 82 msec each
(Scans SRT00501 and SRT00502). There
are four spectra plotted in parallel with time increasing vertically. {\it
Left} The two Parkes polarizations prior to RFI subtraction. {\it Right} The
two Parkes polarizations after RFI subtraction. All spectra are passband
calibrated to compensate for the frequency dependence of the 5 MHz
filters. A third order polynomial spectral baseline was fitted to channels
30-235, 370-405 and 420-500 and subtracted for each time step.}
\label{dynam_acspect.fig}
\end{figure}

Fig~\ref{result.fig} compares uncorrected and corrected spectra for
two scans.
For the purposes of display,
the spectra have been crudely passband calibrated by dividing the
spectra by gain templates formed from the scan average of the total power
spectra of scan SRT00601, which was recorded while the sky frequency for the
Parkes data channels was tuned off the RFI frequency.  Since this gain
template is in common to the processing for both scans SRT00501 and
SRT00502, some of the common structure in the spectra
in Fig~\ref{result.fig}  results from this common gain
template.

The time dependence of the RFI sensed by the Parkes telescope is displayed in
image format in Fig~\ref{dynam_acspect.fig}, side by side with the spectra
after the RFI subtraction.  These spectra received the same
processing as described for the averages in Fig~\ref{result.fig}, with the
additional step of subtracting a third order polynomial spectral baseline
from each time step. This additional step was done to remove some faint
variations in the total power level that occurred from one integration to the
next.  

\begin{figure}[htbp]
\epsscale{1.0}
\plotone{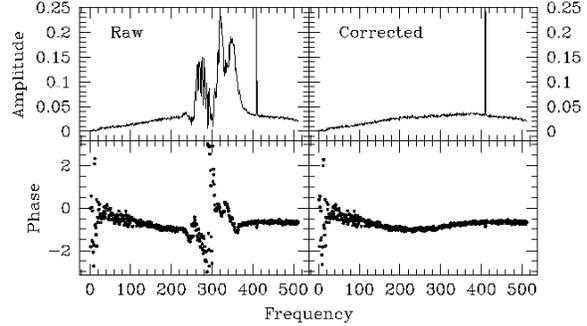}
\caption{Raw and Corrected Parkes A$\times$B
cross power spectra. {\it Left}: The raw, complex cross power spectrum.
{\it Right}: The corrected complex spectrum.}
\label{cross_pwr_AB.fig}
\end{figure}

The cross polarized spectrum from  the Parkes  A$\times$B
is shown in Fig~\ref{cross_pwr_AB.fig}. The figure includes the
raw spectrum
and the corrected spectrum after subtraction of the 
correction shown in Fig~\ref{error_spectra.fig}. 
 Remaining in the corrected 
spectrum, there is a slow modulation of the
power across the 5 MHz band, probably indicating that this power
represents broad band noise
signal that is scattered within the Parkes telescope
structure with the same delay path lengths associated with 
traditional standing waves and path lengths of a few hundred feet.

\begin{figure}
\epsscale{1.0}
\plotone{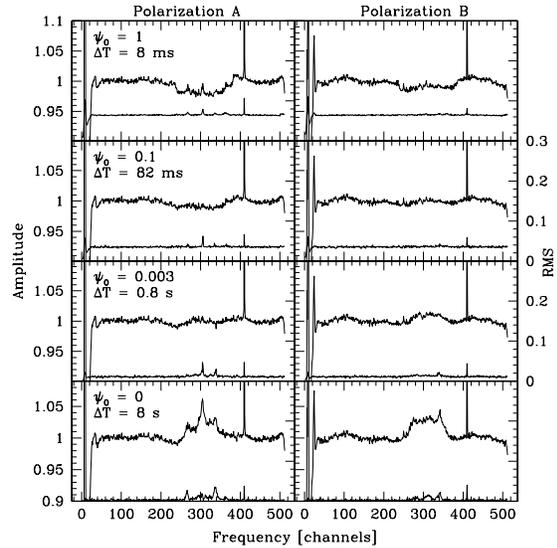}
\caption{The effect of varying the time interval on which
algorithm described in Eqns~\ref{wein1.eqn}-\ref{wein3.eqn}
is applied. The interval varies from 8 msec to 8 sec
in factors of 10.  In each
panel, the upper spectrum shows the  corrected, calibrated, 
baselined spectrum; the lower spectrum is the RMS scatter
about the mean spectrum for each channel as a function
of time. The typical value for RMS should decrease by
$\sqrt{10}$ for each increase in factor 10 in integration
time.
}
\label{amp_rms_vs_inttime.fig}
\end{figure}
 
The integration time over which the RFI corrections are applied is a
critical parameter when the signal to noise ratio of the reference signal
data path is low.  If the integration time is too short, then some of the
derived correction will be noise, and this will be folded into the resulting
spectrum.  On the other hand, if the integration time is long, the impulse
response functions that couple the astronomy and RFI signals to the receiver
will vary, and the mathematics in Eqn~\ref{time_dep.eqn} and
Eqns~\ref{corr1.eqn}-\ref{corr3.eqn} will
break down.  Fig~\ref{amp_rms_vs_inttime.fig} shows a series of tests with a
range of integration times ($t_{int} \sim$~8 msec to 8 sec) for application of the RFI
subtraction.  In all cases, the plots give the grand averages of the entire
25 seconds of the scan after application of the
algorithm on the shorter data segments. The parameter $\psi_o$ 
has an appropriate
value in each case to produce a stable noise level and representation of
spectral features in the frequency range away from the RFI
contamination. For these data, the algorithm was most effective for
integration times of only a $\sim$ 1 sec or less. When treated on 8 sec
averages, substantial RFI remains unsubtracted.

\section{Discussion of Limitations}

The wide range of delays for the scattered RFI led to
problems in our initial experiments, which simulated an FX correlator
with 2048 sample transforms. The RFI illustrated in 
Fig~\ref{raw_cross_power.fig} has two strong components that are delayed
by 15 and 60 time steps. A 60 time step delay is $\sim$3\% of the
time block being processed, so that the cross correlation is not
being performed on fully overlapped data streams. Increasing the
window to 8192 was adequate to reduce the residuals to a level 
that was barely visible above the noise in the corrected spectra.
An additional but less significant improvement resulted from applying
an constant delay offset of 35 time steps to the reference signal
at the input to the ``FX correlator''
to make it closer to the average of the principal delays in 
the Parkes data channels.  A traditional time-domain lag 
correlation spectrometer would not encounter this problem,
provided a sufficient number of lags are allocated to fully
cover the range of delays experienced by the RFI.

In these tests at Parkes, there is a possibility 
that some uncancelled signal may be present due to  a second
transmitter operating at these same frequencies.
The reference horn was pointed at the stronger, nearby transmitter, while the
second transmitter, located at approximately three times the distance, can
be scattered into the Parkes Telescope signal paths without being sensed by
the reference horn.

The coarse digitization of the RFI reference signals will eventually form a
limitation to the precision of the subtraction.  Crude quantization
generates an artificial noise floor throughout the spectrum, effectively
scattering power out of the narrow band RFI.  Since both polarizations from
the reference horn are recorded at relatively high signal to noise ratio,
the quantization noise is also correlated between the two data channels, so
that there is corruption of the cross power spectrum as well as for the
autocorrelation spectra. (The implication is that the term
$<N_3N_4>$ in Eqn~\ref{ccspectra.eqn} will not average to zero with
increased integration.)

\begin{figure}
\epsscale{1.1}
\plotone{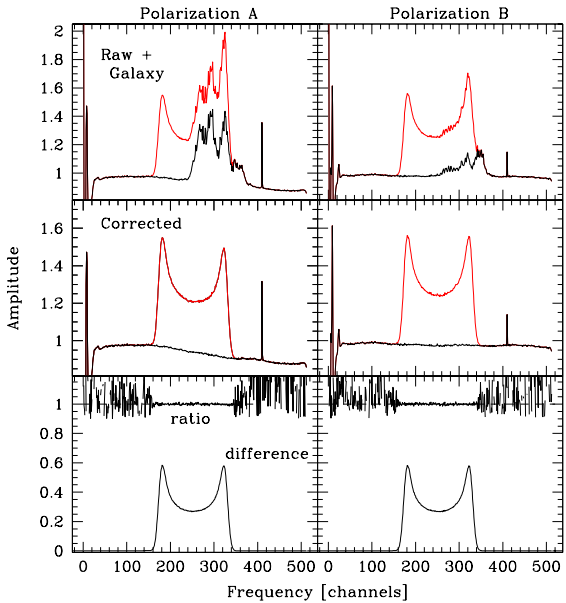}
\caption{Survival of an injected synthetic galaxy signal
in the astronomy channels through the RFI subtraction 
process.
{\it Top}: Raw spectra -- with and without the synthetic
galaxy signal.
{\it Center}: Corrected spectra -- with and without the
synthetic galaxy..
{\it Bottom}: Difference between the corrected spectra
and the ratio between the input
synthetic spectrum and the difference spectrum.
}
\label{gal_tox.fig}
\end{figure}

\section{The Toxicity Test}

 A crucial requirement of an RFI subtraction algorithm is that it
must leave the astronomical signal of interest unaltered.
To test the current method, we added simulated galaxy
signals to the two Parkes input data streams. Independent gaussian
noise was filtered with a double-horned galaxy profile and
the instrumental IF filter passbands and then injected into the
data pipeline just before the correlation stage. Fig~\ref{gal_tox.fig}
shows a comparison between the RFI corrected spectra both with and
without the galaxy.  To highlight the difference between the 
injected signal and the output after RFI subtraction, the bottom
panel shows (1) the difference between the output with and
without the added signal and (2) the ratio of this difference to
the synthetic galaxy profile added to the input. The RMS
deviation of the ratio about unity is 0.005 for polarization A and
0.006 for polarization B.  No systematic deviations are seen across
the band, other than rises in noise level at the edges
where the galaxy profile is approaching zero at the 
edge of the profile. The conclusion is that this method does
no systematic harm to the astronomical signals.

\section{The Parkes Two Feed Experiment}

The mathematical description (Eqns.~\ref{acspectra.eqn} to \ref{wein3.eqn})
 can be equally well applied to a case that uses a second feed from the Parkes
 Telescope as the ``reference horn.'' 
Both feeds may receive astronomical signals, but 
since the feeds point different
 directions in the sky, these are independent astronomical signals, which
will not correlate and therefore will not be subtracted from each
other by this algorithm.  The two feeds  do sense the
same RFI signal $i(t)$, although through different scattering paths.
This is sufficient commonality that the cross power spectral approach
should permit each feed to serve as the reference antenna for the
other.  Similar experiments have  been reported
 \cite{sau97, kse99}.

\begin{figure}[htbp]
\epsscale{1.0}
\plotone{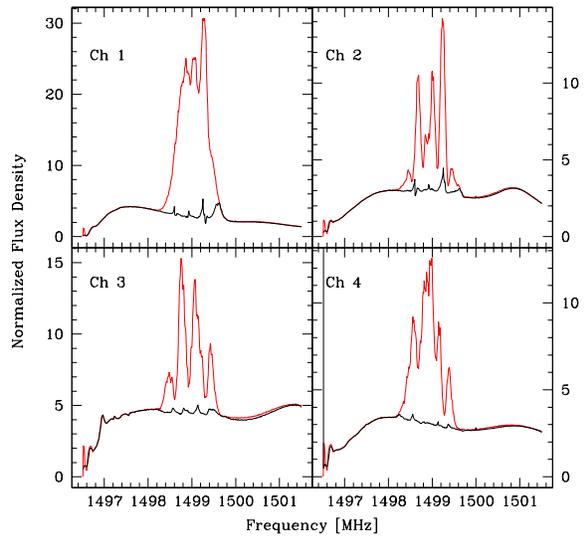}
\caption{The autocorrelation spectra for both polarizations
of two feeds of the
Parkes Multibeam System before and after
rfi subtraction.  An attempt was made to apply a passband
calibration using the same passbands determined from the
scan SRT006\_01 for the reference horn experiment.
The data were treated in $\sim$82~msec averages, which
in turn were integrated over the 20 second duration of
the scan.
}
\label{pks_2feed_ac.fig}
\end{figure}

\begin{figure}[htbp]
\epsscale{1.0}
\plotone{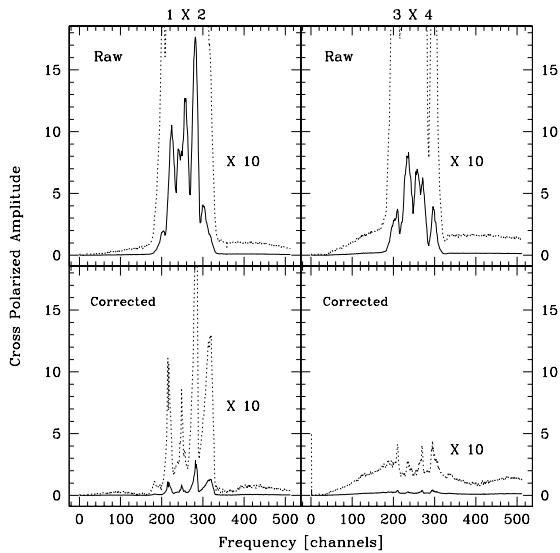}
\caption{The cross polarization spectra for two feeds of the Parkes
Multibeam System before and after rfi subtraction for
scan SRT00108.  An attempt was made to
apply a passband calibration using the same passbands determined from the
scan SRT006\_01 for the reference horn experiment.  The data were treated in
$\sim$82~msec averages, which in turn were integrated over the 20 second
duration of the scan. Dotted lines show the spectra amplified by a factor
10.  }
\label{pks_2feed_2pol.fig}
\end{figure}

Fig.~\ref{pks_2feed_ac.fig} shows a comparison between the autocorrelation
spectra measured for the Parkes two feed experiment, before and after RFI
subtraction.  There is noticeable difference among the four channels
in the effectiveness of the RFI subtraction.  
This probably results from hte bias breated by the broadband polarized
flux or correlated noise in the cross polarized spectra $3{\times}4$
that is being used as the ``reference'' for data channels
1 and 2.  As shown in Fig.~\ref{pks_2feed_2pol.fig},
this noise floor is higher in feed 2 ($INR \sim$ 35:1) than in
feed 1 ($INR \sim$ 100:1), causing the rfi template 
spectra derived from the cross
power spectra  to be less faithful, which in turn
leads to larger error in the correction spectra to be subtracted from 
feed 1. For comparison, the reference horn spectrum $C_{34}(f)$ in
Fig~\ref{raw_cross_power.fig} has $INR \sim$ 1000:1

\section{Conclusions}

RFI subtraction can be performed using cross power spectra between the
astronomy data channels and RFI ``reference'' channels.  In principle, the
reference channel can also be an astronomy channel provide it carries an
astronomy signal that is uncorrelated with the astronomy in the channel that
is being corrected.

The tests made at Parkes demonstrate that a
specifically designed reference sensor provided a higher signal-to-noise
ratio reference signal -- and consequently cleaner cancellation -- than that
obtained from a second horn feed at the Parkes Telescope focus, whose
principal function is to illuminate the Parkes dish.

A refinement will
be to implement this scheme using two reference antennas that are spatially
separated (as in the lower diagram of
Fig~\ref{telescopes.fig}) in order to avoid correlated
noise contributions, while still obtaining as clear and stable path
to the RFI source as possible.  The cross power spectra from the two spatially
separated antennas would form an optimal RFI reference spectrum $C_{34}$
for use in the Eqns~\ref{wein1.eqn}-\ref{wein3.eqn}. To avoid problems
with differential delay causing loss of coherence in the reference signal,
the spectrometer would need to operate with spectral resolution
$\Delta f = (\Delta\phi/2\pi)c/L << c/L$ where $\Delta\phi$ is the allowable
phase rotation across a spectrometer channel and $L$ is the spatial
separation of the sensors. The 2.4~kHz spectrometer resolution emulated in
software for the study reported here would allow spatial separations
of up to 1200m, if $\Delta\phi$ is required to be less than 0.02$\pi$
radians.

There are a number of advantages to performing this type of
``post-correlation'' RFI subtraction: 

\noindent 1. Provided the required
correlation products are recorded (i.e. the online system is 
capable of recording correlation functions with
a sufficiently large number of delay lags), 
the RFI subtraction  can be performed offline, 
where it remains an option in the data reduction path, rather than
a commitment made online and permanently. 

\noindent 2. The method is not vulnerable to the effects of sporadic RFI,
which hurt many algorithms that have an initialization period
while they acquire the RFI signal and optimize their cancellation
parameters.

\noindent 3. Nor is the result influenced by changes in beam shape during
adaptive nulling.

\noindent 4. The correlation method
is effectively a coherent subtraction, since the correlation
functions retain the information describing relative phase between
the RFI entering in the astronomy data stream and the RFI entering
the reference antenna. We showed in section~\ref{accuracy_section}, 
this means that the RFI  
noise power is largely subtracted, leaving only system noise.

\noindent 5. Generalization of the method to an array
of telescopes is straightforward but demands additional
correlator capacity.  If there are two reference signal sensors,
labeled ``$x$'' and ``$y$,''
that sense negligible astronomical signal, then their cross power
spectrum $C_{xy}(f)$ containing a high INR signal
can be used to correct any other power spectrum $C_{ij}(f)$
through the closure relation $C_{ix}^{}C_{jy}^*/C_{xy}^*$.
The $i$, $j$ indices can denote orthogonal or parallel polarizations
drawn from any combinations of antennas in the array 
 or auto correlation, when $i=j$.

\noindent 6. A modification of the method can be applied to
pulsar data streams in which a digital correlator replaces the
narrow band filter bank used in compensating for pulse dispersion.
One possible implementation would 
construct    cross-power ``coupling spectra'' $X(f)$
that are valid for the time interval $t_{int}$
 during which the ``$g$ factors'' of
Eqn~\ref{voltage_spectra.eqn} are stable. The coupling spectrum is
\[
X_{13}(f)=\frac{g_1^*g_4}{g_3^*g_4}\approx\frac{C_{14}^*}{C_{34}^*}
\]
where $C_{14}$ and $C_{34}$ are averaged for up to 1 second as
appropriate. Then the correction $CX_1$ can be computed and applied 
to cancel RFI in measurements of $P_1(f)$ on shorter timescales:
\[
CX_1 = |g_1|^2<|I|^2>=X_{13}g_1g_3^*<|I|^2>\approx X_{13}<S_1S_3^*>
\]
where $<S_1S_3^*>=C_{13}$.
Alternatively one could construct the corrected time
sequence $s_1(t)$ by subtracting the correction
\[
SX_1 = g_1I=X_{13}^*g_3I \approx X_{13}^*S_3
\]
from $S_1(f)$ and inverse Fourier transforming to obtain an RFI
cancelled version of $s_1(t)$.
Computationally efficient schemes could be implemented that include
coherent dedispersion \cite{han74} in the same transform operations
as the RFI cancellation.

\noindent 7. The method can be generalized to removal of 
 solar radiation whose multi-path scattering effects
give rise to the spectral ``standing wave'' problem.  The
important difference that the Sun generates a dual polarized signal
with a variable polarized component; this will necessitate a processing
path more akin to the sub-space decomposition \cite{ell99,lvb00}, 
in order to identify orthogonal components of the solar ``rfi'' signals.

\noindent 8. The present generation of digital correlators, which
typically operate with single bit to 9 level precision, could
implement this method for use in moderate levels of RFI for testing
and astronomical observation in the near future.

The disadvantages of the method are: (1) The data rates will be high, since the
method requires a full-scale cross correlator, preferably with multi-bit
precision to accept large SNR RFI reference signals, that must dump spectra
after relatively short integrations of less than $\sim$1~sec.  Of course,
these data rates are lower than recording the full baseband, but they are
substantially higher than a real-time adaptive filter approach, which would
allow long spectral integrations once the RFI has been canceled. 
(2) When the interference is much stronger than the astronomical signal,
a 1 or 2 bit sampler is ``captured'' so that the data stream consists
primarily of $\pm$2, and the zero crossings are determined by the phase of
the interferer. Under these conditions, the astronomical data will be
largely lost. These applications will require correlators with greater
digital precision.

The case where multiple interferers occupy the same frequency band
will require a greater number of sensors and more correlator
capacity devoted to processing the the RFI signals, as laid out in the
analyses of Sault (1997) \nocite{sau97} and Ellingson (1999) \nocite{ell99}.

\acknowledgments

The authors are grateful to W. van Stratten, M. Bailes, S. Anderson,
S. Ellingson, R. Sault, P. Perillat,
 R. Ekers, J. Bunton, L. Kewley, M. Smith and P. Sackett for
helpful comments and discussion.  F.B. is grateful to the ATNF in Epping,
NSW, the Department of Astronomy at OSU, Columbus, OH,
and to the IAS, Princeton, NJ for their hospitality while this
work was done.


\end{document}